# Harnessing Single Polarization Doppler Radars for Tracking Desert Locust Swarms

N. A. Anjita[1], J. Indu[1,2], P. Thiruvengadam[3], Vishal Dixit[2], Arpita Rastogi[4] and Bagavath Singh Arul Malar Kannan[4]

[1] *Department of Civil Engineering, Indian Institute of Technology, Bombay, Powai, Mumbai, India*

[2] *Interdisciplinary Programme in Climate Studies, Indian Institute of Technology Bombay, Powai, Mumbai, India*

[3] *School of Meteorology, University of Oklahoma, Norman, Oklahoma, United States.*

[4] *India Meteorological Department, New Delhi-110003*

*Corresponding Author: J. Indu; indusj@civil.iitb.ac.in*

## Abstract

Desert locusts, notorious for their ruinous impact on agriculture, threaten over 20% of Earth's landmass, prompting billions in losses and global food scarcity concerns. With billions of these locusts invading agrarian lands, this is no longer a thing of the past. Recent invasions, such as those in India, where losses reached US $3 billion in 2019-20 alone, underscore the urgency of action. By tapping into the existing Doppler Weather Radar (DWR) infrastructure, originally deployed for meteorological applications, the present study demonstrates a systematic approach to distinctly identify and track concentrations of desert locust swarms in near real-time using single-polarization radars. Findings reveal the potential to establish early warning systems with lead times of around 7 hours and spatial coverage of approximately 100 kilometres, empowering timely mitigation efforts. In stark contrast to satellite imagery, often constrained by resolution limitations that impede swarm detection, DWRs offer unparalleled spatial and temporal resolution, rendering them indispensable tools in this endeavour. Harnessing radar capabilities within the current infrastructure, unleashes unparalleled ecological monitoring potential and fundamentally transform practices in managing migratory pest management practices. Embracing these technological advancements becomes imperative to safeguard agricultural landscapes, uphold global food security, and effectively mitigate the ecological threats posed by migratory pests.

***Keywords***: *Aeroecology, Agricultural Pest, Desert Locust, DWR, Entomology,*

# 1. Introduction

Desert locusts (Schistocerca gregaria) are voracious pests known for their ability to form massive swarms, each consisting of millions of individuals. They unleash havoc on agricultural landscapes, resulting in widespread crop devastation and posing a significant threat to food security. Desert locusts initially manifest as solitary individuals dwelling in arid desert regions. When favourable environmental conditions materialise, they congregate, reproduce prolifically, and coalesce into formidable swarms. These swarms embark on migratory journeys, traversing from one region to another in pursuit of sustenance *(Cressman, 2016; WMO & FAO, 2016)*. Left unmitigated, their uncontrolled movement can potentially unleash global havoc in the agricultural sector. In India, a nation heavily reliant on agriculture, the desert locust upsurge during 2019-20, as analysed by *Chatterjee et al. (2020)*, led to estimated total losses of approximately US$ 3 billion. This sum accounted for 13% of the overall funds allocated by the Indian government for agriculture, irrigation, and related activities. These statistics highlight the urgent necessity to monitor and mitigate locust outbreaks.

Many countries rely on collaboration between radar meteorologists and entomologists to track migratory insects in the troposphere to achieve a better process-level understanding of atmospheric dynamics (*Rennie et al., 2014*; *Cui et al., 2020; Gauthreaux & Diehl, 2020)*. This entire insect tracking or weather prediction gamut relies on dense observational networks to provide improved forecast services. Weather radars represent one such vast observational network, immensely contributing to the operational forecasting systems of various countries by regularly sweeping the airspace for both meteorological and entomological applications. Through this study, we leverage an existing Doppler Weather Radar (DWR) network and geospatial tools to detect and track desert locust swarms across the agrarian regions of central and north-northwestern India (Fig. 7).

Over the years, monitoring desert locusts has relied on remote sensing-derived meteorological and ecological variables such as precipitation, vegetation, soil moisture, air temperature and soil temperature *(Latchininsky & Sivanpillai, 2010; Cressman, 2013; Escorihuela et al., 2018; Piou et al., 2019; Gómez et al., 2020)*. Recently, UAV-based technologies have also been deployed to identify areas suitable for oviposition and track locust nymph bands *(Klein et al., 2021)*. However, direct monitoring of locust populations during flight poses significant challenges, mainly owing to the scarcity of exceptionally high-resolution and readily accessible data on an operational basis *(Cressman, 2013)*. This is where the existing countrywide DWR networks can prove to be an indispensable asset in leveraging our understanding of migratory insects like desert locust swarms (*Drake & Reynolds, 2012*).

DWRs, designed for meteorological applications, have been extensively used to detect and study biological targets such as birds, bats, and insects for more than 40 years. Mainly, the S-band (2700–2900 MHz), C-band (primarily 5300–5650 MHz), and X-band (9300–9500 MHz) are the commonly used frequency bands, playing a pivotal role in weather radar applications. While entomological radars are adept at measuring individual insect speed, the direction of movement, orientation, shape, and sometimes wingbeat frequency *(Drake, 2016)*, DWRs have their unique advantage in scanning horizontally and at different elevations. This enables them to play a crucial role in detecting locust swarms during active migration and studying their vertical structures. Desert locust swarms exhibit distinctive characteristics in size, density, composition, and opacity, setting them apart from other radar targets. Nevertheless, they contain valuable information often discarded as clutter in meteorology. Notably, in radar meteorology, locust swarms are considered as one of the factors causing clear air (rain-free) echoes or 'angels.' In entomological applications of DWR, the targets seldom represent single insects; instead, they typically depict concentrations or volumes of insects. Traditionally, the altitudinal distribution of desert locusts is highly stratified, with dense layers forming at varying altitudes, typically 50-150 meters deep *(Schaefer, 1976)*. Their flight generally occurs at altitudes below 2 kilometres *(Cressman, 2016)* in the troposphere, which also houses part of the weather phenomena. Globally, numerous nations maintain a robust network of DWRs for comprehensive weather monitoring and forecasting *(Saltikoff et al., 2019)*. Australia serves as an example, with DWRs strategically positioned nationwide for extensive weather monitoring, which has also been employed to detect and study patterns of Australian Plague Locust populations and migrations. Similarly, the impressive NEXRAD network in the US, consisting of around 260 DWRs, is being utilised for meteorological and aeroecological applications. The recession and invasion zones of desert locusts *(Cressman, 2016; Anjita & Indu, 2023)* reveal significant outbreaks occurring in Asia, particularly in central and north-northwestern India, between 2019 and 2021. Notably, these locations also house DWRs scanning regions predominantly of agrarian population.

Understanding when and where locust swarms terminate their migration can be crucial for directing mitigating actions. Despite the importance of understanding locust trajectories at large spatial and temporal scales, very few studies are available in India. The dearth of studies is primarily due to limitations in accessing suitable data. Leveraging the existing India Meteorological Department (IMD) weather radar network, we harnessed the locust tracking abilities of existing single-polarization DWRs to provide the first view of locust tracks. Additionally, we created maps detailing locust trajectory and velocity of migration towards the predominantly agrarian regions of central and north-northwestern India.

## 2. Results and Discussions

For this analysis, we processed 133 volume scans from a single-polarized DWR in Lucknow. The primary variables for the present study included reflectivity (Z), spectrum width (W) and radial velocity (V). Following a preliminary visual inspection of the data, it was observed that significant signatures were lacking in the variable, W. Consequently, this was excluded from the subsequent filtering methodology. The scanning strategies entail automatically elevating the radar antenna to progressively higher pre-set angles during its rotation. As the radar traverses through all elevation angles, a complete volume scan of the atmosphere is accomplished. The compilation of elevation scans, or slices, constitutes a Volume Coverage Pattern (VCP). The flight altitude of desert locusts is considered to be below 2 km. This is mainly because they are known to fly at the level of the warmest air, which is best suited for migration, rather than at higher altitudes with suboptimal lower temperatures. Considering the probable altitudinal distributions, a threshold height of 2 km is employed to exclude all gates (also referred to as bins or pixels) in the radar images during the lowest elevation scans, with an elevation angle of approximately 0.2° (for S-band). This restricts the radar coverage for analysis to 150 kilometres (Fig. 1).

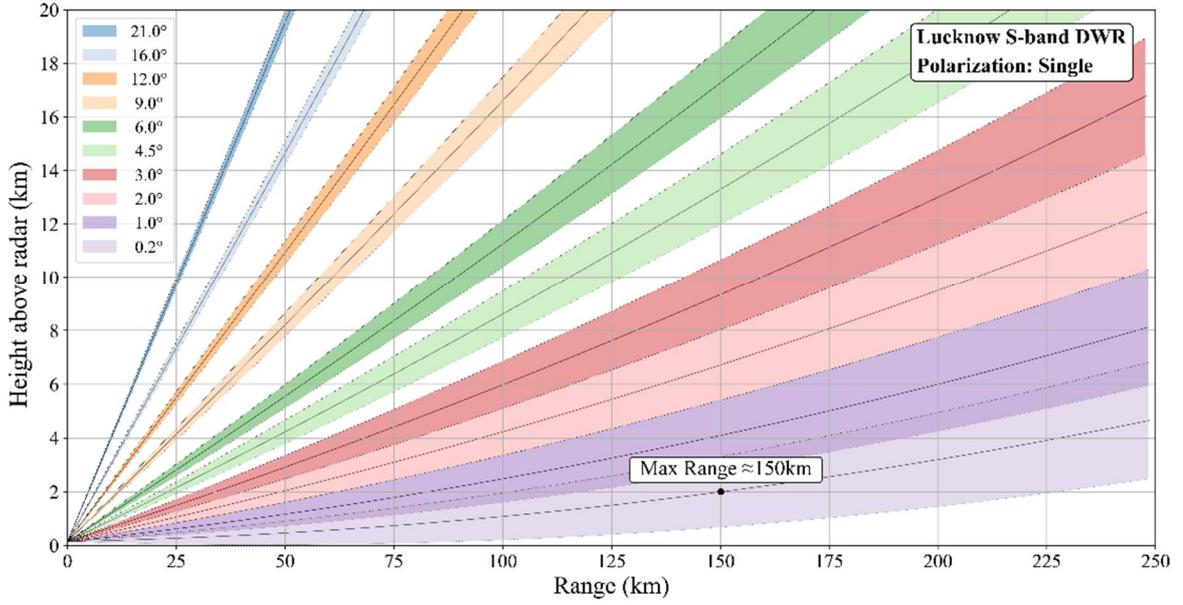

*Fig. 1: Volume Coverage Patterns (VCPs) for the Lucknow S-band DWR*

For DWRs operating at C and S-band frequencies, locust swarms are seen as Rayleigh targets (*Drake & Reynolds, 2012*). Surveillance using DWR scans helps qualitatively examine the development and decay of the swarm concentrations and their movements across the landscape. As the initial step, a fully automated quality check and analysis involving filtering the echoes from lowest elevation scans at both sites, focusing on the variables Z and V is conducted. Filtering was based on radial velocity, V and reflectivity, Z. As radial velocities will always be lesser than or equal to actual target velocities, setting a threshold of 6m/s for V effectively removed echoes moving at speeds beyond this limit. This decision aligns with the known flying speed of locust swarms, estimated to be around 16 to 19 km/hr *(or 4.4 to 5.3 m/s) (FAO)*. The Z threshold was set at +15 dBZ, removing all signals below this level, determined qualitatively through a detailed examination of the DWR images with the swarm echoes and referencing previous literature (*Amarjyothi et al., 2022; Gauthreaux & Diehl, 2020*). This choice is supported by *Rainey's (1955)* study, suggesting that desert locust swarms can produce high Z values comparable to those observed in precipitation.

However, for the single-polarization DWR, the absence of dual-pol capabilities, especially RHOHV, prevents the effective filtering of precipitation echoes and they can also result in retaining echoes from convective storms, which in northeastern India during the monsoon season typically move at speeds between 3-4 m/s. Additionally, shallow convective storms occur below 5 km altitude *(Jha et al., 2023)*. Hence, the retained echoes can only be confirmed as those of desert locust swarms by checking for precipitation on that day around the DWR site and cross-checking with news reports or alerts.

Groups with fewer than five pixels (with gate sizes of 250 m for Lucknow S-band DWR), which are too small to represent a swarm, were eliminated, further refining the data. The resulting surveillance observations in Fig. 2 display the filtered echoes of the desert locust swarm within the radar image.

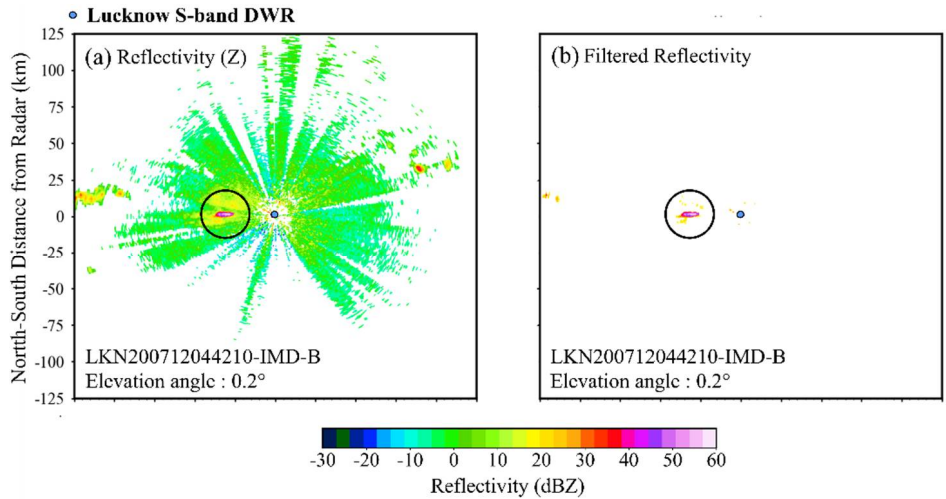

*Fig. 2: (a) depicts the raw reflectivity scans obtained from the Lucknow S-band DWR on July 12, 2020, at 03:52:10 UTC. In contrast, (b) depicts these scans with only the echoes generated by desert locust swarms preserved while others are removed.*

Post filtering, the contiguous groups of pixels depicting desert locust swarm echoes were analysed based on their average reflectivity and pixel count. In Lucknow, the average pixel count was 2880, with an average reflectivity of 27.11 dBZ. These findings are further elaborated in Section 2.1.

## 2.1 Demarcating Biological and Meteorological Echoes

As continued observations of dynamic features help track the target development, a long-time series of locust activity was obtained for this study. Radar observations of locust swarms rely on the back-scattering of radio waves off them. As such, a single numerical value cannot fully represent the reflecting ability of locust swarms. This is due to multiple factors, such as the wavelength, polarization and the direction the locusts face relative to the radar beam. While operating in the C and S-band frequencies, insects such as locusts are observed as Rayleigh targets. Due to their non-spherical shape, the radar cross-section of these targets, which describes their effectiveness at reflecting a signal back to the radar, also fluctuates with changes in aspect, polarization and roll angles (*Drake & Reynolds, 2012*). Moreover, the radar reflectivity factor also gets modulated owing to wing beating, which further complicates the interpretation of echoes recorded by a DWR.

As the DWR scanning strategy involves the radar antenna covering a wide area horizontally and over different elevations, the layers of insect concentrations shall be clearly revealed and can be demarcated. To check the same, we analysed the vertical reflectivity plots from both the DWR sites along a latitude intersecting the identified locust swarm echo. Fig. 3 present side-by-side composite reflectivity plots and vertical cross-sections of reflectivity at different timestamps. These cross-sections are displayed along a selected latitude to visualise the vertical structure of the echoes produced by the swarms. The horizontal coverage is restricted to 150 km in both the eastward and westward directions from the Lucknow DWR, which were values initially determined as the maximum range within which signals from the swarm would be observed, based on the maximum flying altitude of locusts. In Fig. 3, depicting the Lucknow images, a bright cluster is seen slowly shifting towards the radar centre. Upon closer inspection of the vertical reflectivity cross-sections on the right, a continuous shift of these echoes over time is observed, consistently remaining below the 2 km threshold initially set. The reflectivity cross-sections consistently indicate that the height of these echoes never exceeds 2 km throughout the movement of the desert locust swarm. This validation serves as a critical confirmation of both the extraction methodology and the resulting findings.

The observed average pixel count for swarms by the S-band DWR, at 2880, appears disproportionately high, potentially due to the retention of both desert locust swarms and non-precipitation returns. The discrepancy arises due to the limitations of the single-polarized S-band DWR, which cannot discern the shape of individual targets, unlike dual-polarized systems. This functionality provided by dual-polarized systems is crucial for distinguishing between biological and meteorological echoes. Meteorological radars are increasingly integrating dual-polarization technology, as polarimetric measures such as RHOHV aid in distinguishing between various targets and provide a wealth of additional information about them.

To further validate that the extracted echoes are indeed from desert locust swarms, we meticulously cross-referenced data with precipitation records obtained from the AWS (Automatic Weather Stations) - ARG (Automatic Raingauge Stations) networks, IMD (Indian Meteorological Department) available at *http://aws.imd.gov.in:8091/*. We pinpointed AWS stations within a 150 km radius of the Lucknow DWR, making sure to select those with available data. For the scenario on July 12$^{th}$, 2020 (from 02:00:00 to 11:00:00 UTC), stations selected in Uttar Pradesh were Auraiya, Kanpur, Lucknow, Fursatganj, Sultanpur, and Unnao AMFU. Notably, none of these stations recorded any rainfall on the specified dates. This supports the conclusion that the continuous high reflectivity echoes observed on the radars are unequivocally linked to desert locust swarms. Moreover, news reports also corroborate the presence of locust swarms in these locations on the specified dates.

Through the swarm echoes from both DWRs, we analyse the parameters of biological interest, namely timing, height, and flight direction. It is important to note that identification is achieved for a population (i.e., insect concentrations) and not at an individual level. With local survey information, these parameters can be further linked to specific species of locusts predominant in the area. Consequently, DWR echoes can be discernible to locust swarms even without an expert taxonomist, supported by supplementary evidence.

*Fig. 3: Visual representation of desert locust swarm echoes below 2 km threshold: Lucknow DWR*

**2.2 Desert Locust Swarm Dynamics**

As a next step, the study explores the speed of swarm movement and the distance covered and investigates whether the locust swarms exhibit a mind of their own or if wind patterns solely influence their flight path and direction. To explore this, swarm-returned echoes are meticulously traced as long as they remain within the radar coverage.

If the speed at which the locust swarms are migrating and the flight duration are known, this translates to the distance traversed by the swarm. For instance, in the case of desert locust swarm echoes observed on the Lucknow DWR on 12$^{th}$ July 2020, starting at 02:32:10 UTC and ending at 11:32:09 UTC, we estimated the displacement run to be 108.87 kilometres. The swarm progressed at an average velocity rate of 3.47 m/s. Further, we utilised ERA5 hourly wind speed data (u- and v-components of wind ms$^{-1}$) at appropriate pressure levels *(Hersbach et al., 2023) (obtained from the Climate Data Store (CDS))* to examine wind patterns and magnitude in the vicinity of the DWR stations (Fig. 4). Wind speeds and directions were charted for both the onset and cessation of the swarms on the radar. Streamlines were also plotted around the radar station to observe the alignment of swarm movements with the wind. The winds are known to have scale-dependent behaviour. In this scenario, as observed in Lucknow, swarm movements were predominantly aligned with the broader-scale (25 km) wind patterns, with minimal deviation, generally following the downwind direction. Nonetheless, wind alone may not be an exclusive factor influencing their flight path *(Sorel et al., 2024)*. Multiple other factors, including humidity, temperature or food availability, may influence flight patterns. Hence, it is crucial to emphasise that further research is necessary to substantiate these observations, and such investigations will be a focal point in future studies.

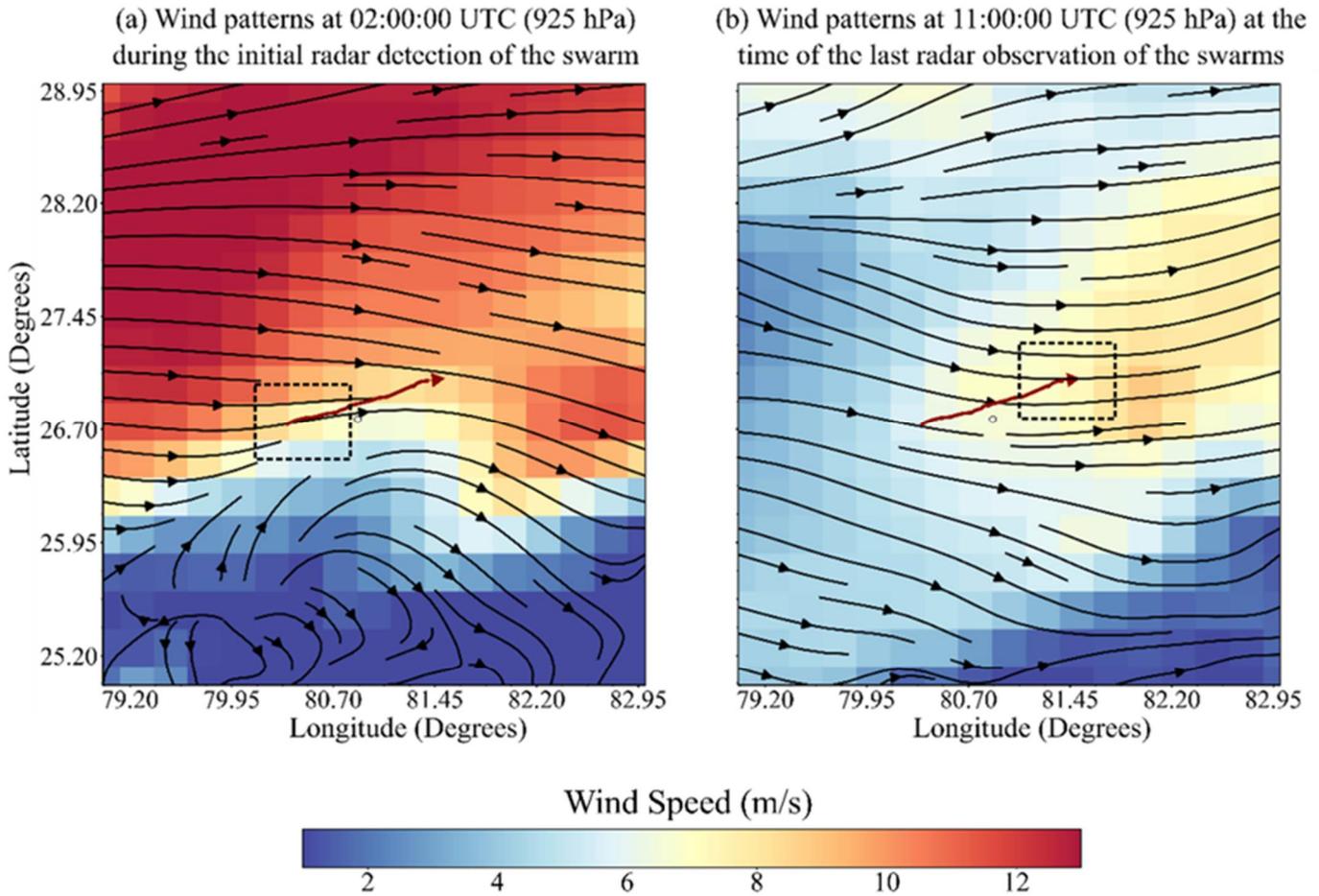

*Fig. 4: Wind patterns surrounding DWR station and desert locust swarm trajectories*

## 3. Opportunities

In summary, by taking advantage of an existing weather radar infrastructure, the present study provides baseline evidence on passages of locust swarms across central and north-northwestern India. As surveillance targets of locust concentrations can be plumes of insects rising in convective updraughts or linear accumulations at weather fronts, there is a need to delineate meteorological echoes, especially those from desert locust swarms. Also, the study's findings present qualitative evidence to establish and track radar echoes of desert locust swarms.

Even with limited high-tech resources within the recession-invasion limits of desert locusts, we can still tap into and utilize the existing infrastructures to track swarms across borders. Further, single-polarized DWRs, having the ability to detect swarms, facilitate their tracking and enable the implementation of cost-effective early warning systems for desert locusts with minimal additional expenses. In essence, this study serves as a clarion call for the widespread global adoption of radar aeroecology methodologies, advocating the untapped potential of existing weather radar network infrastructures for ecological monitoring.

Further, understanding the dynamics of desert locust swarms, their current locations, and forecasted movements is crucial for authorities to take timely and effective measures to mitigate the devastating impact on crops and livelihoods. With continued research, the primary goal remains to establish an operational near-real-time tracking and forecasting system for locust swarms. This endeavour holds immense promise for profoundly aiding severely affected countries, empowering them to proactively address disasters and shield their agricultural landscapes and communities from devastation.

## 4. Materials and Methods

### 4.1 DWR Data Download

The IMD is the authoritative body overseeing the upkeep and functionality of DWRs, which are strategically positioned throughout India. The DWR network in India comprises 5 C-band, 22 S-band and 12 X-band DWR (*https://mausam.imd.gov.in/*) (Fig. 5). Each DWR collects azimuthal scans of the atmosphere every 10 minutes at ten elevation angles: 0.2°, 1.0°, 2.0°, 3.0°, 4.5°, 6.0°, 9.0°, 12.0°, 16.0°, and 21.0° for Lucknow *(Roy Bhowmik et al., 2011)*. Considering the geographical boundaries of desert locust invasion and recession areas and reports of past locust attacks over India *(Press Information Bureau, Govt. of India)*, a DWR actively operating in these limits over India was chosen for the study (Lucknow). This study centres on the S-band (single-polarized) deployed at these sites. The radar scans of July 2020 for Lucknow were identified as crucial periods due to the presence of desert locust swarm sightings in the region during this time *(Amarjyothi et al., 2022)*. In this study, Python ARM Radar Toolkit (Py-ART) *(Helmus & Collis, 2016)*, wradlib *(Heistermann et al., 2013)* and PyScanCf *(Syed et al., 2021)* were were utilized to generate various radar plots.

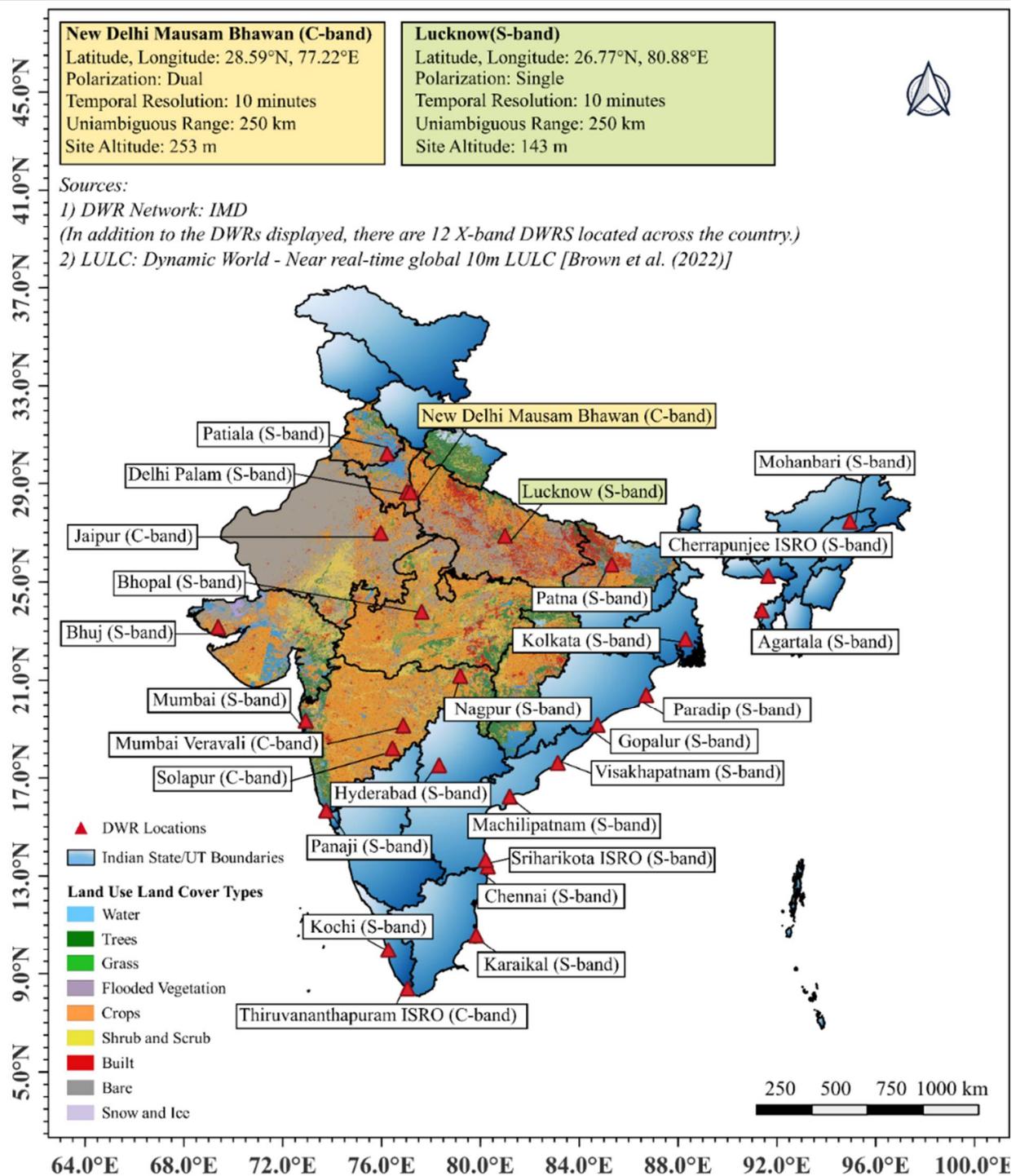

Fig. 5: India Meteorological Department (IMD) Doppler Weather Radar (DWR) Network (C- & S-band), highlighting the Lucknow single-polarization radar and the New Delhi (Mausam Bhawan) dual-polarization radar. Also depicted are the land use and land cover (LULC) of states impacted by desert locust swarms during 2019-2021. The LULC maps are generated using the Dynamic World (Near real-time global 10m) LULC dataset (Brown et al., 2022) in Google Earth Engine and imported into QGIS 3.22 Białowieża. In QGIS, these maps are visualized along with the IMD DWR network (Source: IMD).

**4.2 Adjusting Range Limits**

Depending on a selected VCP, DWRs accomplish a full volume scan approximately every 5–10 minutes, presenting a detailed depiction of the surrounding atmosphere *(Cui et al., 2020)*. The VCPs were visualised for each station using equations (1) and (2) *(George et al., 2011)*.

$$Height, h = \sqrt{\left(S^2 + \left(\frac{4}{3} \times R\right)^2 + 2 \times S \times \frac{4}{3} \times R \times \sin(\theta)\right)} - \left(\left(\frac{4}{3} \times R\right) + \left(\frac{h_r}{1000}\right)\right) \quad (1)$$

$$Range, S = S_1 + \Delta S \times n \quad (2)$$

Where 'h' represents the altitude of the observation point above mean sea level in kilometres, 'θ' stands for the elevation angle measured in radians, '$h_r$' signifies the radar height in meters, and 'R' denotes the radius of the earth (6371 km). The range 'S' is calculated using details from the distance to the first range bin ($S_1$) and the interval between two range bins (ΔS) specified in the netCDF file's header. The variable 'n' assumes 1, 2, 3, 4, and so forth, corresponding to specific bin values *(George et al., 2011)*.

The initial filtering accounts for the maximum probable flying height of desert locust swarms, established as 2 km above the ground based on past literature *(Amarjyothi et al., 2022; Cressman, 2016; Drake & Farrow, 1983)*. We calculate the maximum observable range within which a swarm can be detected and confine further analysis within this range.

**4.3 Radar Filtering**

For each scan, we first worked to remove the meteorological echo by applying thresholds to reflectivity, radial velocity and spectrum width. Echoes are then grouped based on their interconnectedness *(Stepanian et al., 2014),* resulting in discernible pixel groups. Groups with fewer than five pixels are deemed too small to constitute a swarm and are removed to finally retain pixels belonging to the desert locust swarm. The retained echoes attributed to the swarms are characterised based on their average reflectivity, pixel count, velocity, and distance travelled.

**Data Availability**

The Doppler weather radar data that support the findings of this study are available from the India Meteorological Department (IMD) but restrictions apply to the availability of these data, which were used under license [MOES/REACHOUT/CAN/2022] for the current study, and so are not publicly available. The Dynamic World V1 LULC dataset and MODIS NDVI data can be accessed through Google Earth Engine. ERA5 data can be obtained from the Copernicus Climate Data Store (https://cds.climate.copernicus.eu/cdsapp#!/dataset/reanalysis-era5-pressure-levels?tab=form) .## Acknowledgement

The authors thank the support of the Ministry of Earth Sciences through the research grant MOES/REACHOUT/CAN/2022. The authors acknowledge the support by India Meteorological Department (IMD) and the facilities extended by IIT Bombay.## Ethics declarations

### Competing interests

The author(s) declare no competing interests.

18